\documentclass[aps,prl,twocolumn,superscriptaddress,groupedaddress]{revtex4-1}  
\usepackage{graphicx,amssymb}
\usepackage{color,ulem}
\usepackage{bm}

\newcommand{\eqref}[1]{(\ref{#1})}

\begin{document}
\title{Correlated vs. uncorrelated noise acting on a quantum refrigerator}

\author{Bayan Karimi}
\affiliation{Low Temperature Laboratory, Department of Applied Physics, Aalto University School of Science, P.O. Box 13500, 00076 Aalto, Finland}
\author{Jukka P. Pekola}
\affiliation{Low Temperature Laboratory, Department of Applied Physics, Aalto University School of Science, P.O. Box 13500, 00076 Aalto, Finland}

\date{\today}

\begin{abstract}
Two qubits form a quantum four-level system. The golden-rule based transition rates between these states are determined by the coupling of the qubits to noise sources. We demonstrate that depending on whether the noise acting on the two qubits is correlated or not, these transitions are governed by different selection rules. In particular, we find that for fully correlated or anticorrelated noise, there is a protected state, and the dynamics of the system depends then on its initialization. For nearly (anti)correlated noise, there is a long time scale determining the temporal evolution of the qubits. We apply our results to a quantum Otto refrigerator based on two qubits coupled to hot and cold baths. Even in the case when the two qubits do not interact with each other, the cooling power of the refrigerator does not scale with the number ($=2$ here) of the qubits when there is strong correlation of noise acting on them.
\end{abstract}


\maketitle
Controlling the susceptibility of qubits to decoherence sources is a central issue in developing a robust quantum computer \cite{clarke, devoret, nielsen, governale, storcz, pashkin,averin,you, hu}. Over the past two decades it has become obvious that the influence of a common source of noise on all qubits deviates dramatically from the situation where uncorrelated noise sources affect each individual qubit separately \cite{zanardi, d'arrigo, gustavsson, galperin}. In the first case, so called decoherence-free subspaces emerge, meaning that there are states that are not affected by the noise source. To realize robust quantum circuits and to provide error correction \cite{barends} for them depends then on whether the noise sources are correlated or not \cite{averin1, clemens, klesse, novais}.

\begin{figure}[t]
\centering
\includegraphics [width=0.9\columnwidth] {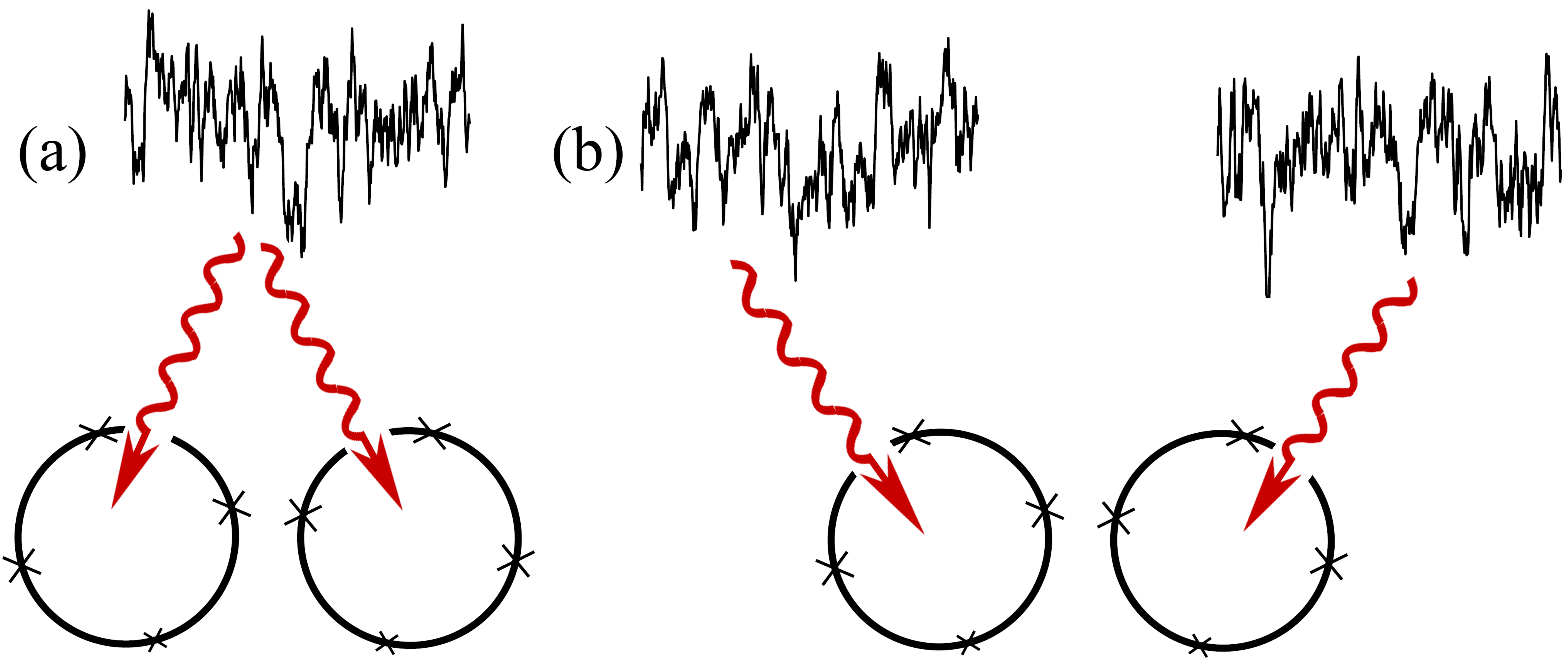}
\caption{Two qubits subjected to (a) correlated and (b) uncorrelated noise sources. 
\label{fig1}}
\end{figure}

In this Letter we demonstrate selection rules that account for the transitions in a two-qubit system affected by either a common noise source or multiple sources shown schematically for two extreme cases in Fig. \ref{fig1}. The basic four-level system of the two qubits exhibits then a protected state ("decoherence-free subspace") when subjected to fully (anti)correlated noise, whereas for uncorrelated noise, all the four states couple to the noise. In our work we focus, instead of qubit decoherence, on energy transport between the baths at different temperatures producing the noise on the qubits. We study the dependence of this transmitted power on the initialization of the system and on the degree of noise correlation. To understand the influence of noise correlation in a physical system, we investigate a quantum Otto refrigerator \cite{Jukka, niskanen2007}, a representative of quantum machines that are currently of considerable interest due to their experimental feasibility \cite{kay, campisi2016, hofer2016, quan2007, rossnagel2016, Uzdin/Kosloff, brandner2016, niskanen2007, Jukka}. 

\begin{figure*}[t!]
\centering
\includegraphics [width=18cm] {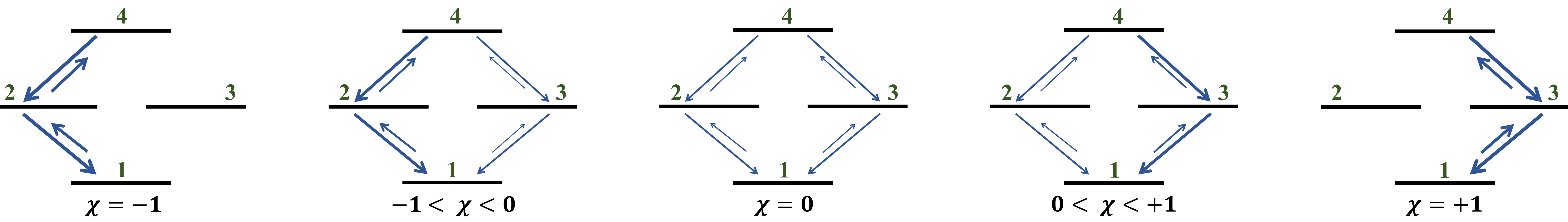}
\caption{a) Transition rates in the four level system of decoupled qubits for different levels of noise correlation $\chi$.
\label{fig2}}
\end{figure*}
The total Hamiltonian describing the system and the environment is given by
\begin{equation} \label{hamiltonian}
H =H_{\rm Q1}+H_{\rm Q2}+ H_{\rm S}+H_{\rm cS,1}+H_{\rm cS,2},
\end{equation}
where $H_{\rm Q1}, H_{\rm Q2}$ are the Hamiltonians of the two (driven) qubits, $H_{\rm S}$ is the Hamiltonian of the noise source(s), and $H_{\rm cS,1}, H_{\rm cS,2}$ the couplings of qubits 1 and 2 to the noise source(s). For our main argument we may assume that the two qubits are mutually decoupled although the selection rules to be presented hold also for coupled qubits. In the quantitative analysis, we use the four Bell basis states $\{|u_1\rangle=\frac{1}{\sqrt{2}}(|0_10_2\rangle +|1_11_2\rangle),~|u_2\rangle=\frac{1}{\sqrt{2}}(|0_10_2\rangle -|1_11_2\rangle),~|u_3\rangle=\frac{1}{\sqrt{2}}(|0_11_2\rangle+ |1_10_2\rangle),~|u_4\rangle=\frac{1}{\sqrt{2}}(|0_11_2\rangle- |1_10_2\rangle)\}$. Here the subscript $j=1,2$ on the rhs refers to the qubit $j$, for which
\begin{equation} \label{s1}
H_{{\rm Q}j}= - E_j(\Delta_j \sigma_{x,j} + q\sigma_{z,j}),
\end{equation}
with $E_j$ the overall energy scale of each qubit, $ \sigma_{x}$ and $\sigma_{z}$ the Pauli matrices, $2\Delta_j$ the dimensionless energy splitting at $q=0$, and $q$  is the flux applied equally to both qubits. We assume that the system is fully symmetric, i.e. $E_0\equiv E_1=E_2$, and $\Delta \equiv \Delta_1=\Delta_2$ and that all the noise sources and their couplings to the individual qubits are equal. The eigenenergies of the Hamiltonian (in units of $E_0$) are given by
\begin{equation} \label{s2}
\lambda_1=-2\sqrt{q^2+\Delta^2},~~ \lambda_2=\lambda_3=0,~~\lambda_4=+2\sqrt{q^2+\Delta^2}
\end{equation}
and the corresponding eigenstates are
\begin{eqnarray} \label{s3}
&&|1\rangle=\frac{1}{\sqrt{2}}(|u_1\rangle+\frac{q}{\sqrt{q^2+\Delta^2}}|u_2\rangle+\frac{\Delta}{\sqrt{q^2+\Delta^2}}|u_3\rangle)\nonumber\\&&
|2\rangle=|u_4\rangle\nonumber\\&&
|3\rangle=\frac{\Delta}{\sqrt{q^2+\Delta^2}}|u_2\rangle-\frac{q}{\sqrt{q^2+\Delta^2}}|u_3\rangle\\&&
|4\rangle=\frac{1}{\sqrt{2}}(|u_1\rangle-\frac{q}{\sqrt{q^2+\Delta^2}}|u_2\rangle-\frac{\Delta}{\sqrt{q^2+\Delta^2}}|u_3\rangle).\nonumber
\end{eqnarray}

For the noise, we consider a generic form of linear coupling between each qubit and the noise source as 
\begin{equation}
H_{\rm cS}\equiv \sum_{m=1,2} H_{{\rm cS},m} =\sum_{m=1,2} \hat{A}_m\delta\hat{X}_m(t),
\end{equation}
where $\hat{A}_m$ determines the coupling and $\delta\hat{X}_m(t)$ is the time $t$ dependent fluctuation of the quantity. In what follows we investigate the cases of different degrees of correlation between two noise fluctuators. The noise correlators with the help of their Fourier transform are given by
\begin{equation}\label{correlator}
\langle \delta\hat{X}_m(t')\delta\hat{X}_n(t)\rangle=\chi_{mn}\int_{-\infty}^{\infty} \frac{d\omega}{2\pi}e^{-i\omega(t'-t)}S(\omega),
\end{equation}
where $S(\omega)$ is the unsymmetrized noise spectrum of each current. Here, we define $\chi_{mn}$ as the degree of correlation of noise sources $m$ and $n$. For autocorrelation, $\chi_{11}=\chi_{22}=+1$, and for crosscorrelation we set $\chi_{12}=\chi_{21}=\chi$, where $-1\le \chi\le +1$. If $\chi=+1$, noise is fully correlated and for $\chi=-1$ anticorrelated, whereas for $\chi=0$ we have uncorrelated noise from independent sources. 

The transition rates from the $k^{th}$ to the $l^{th}$ instantaneous eigenstate due to the noise source(s) can be calculated from the Fermi's golden rule as
\begin{equation}
\Gamma_{k\rightarrow l,{\rm S}}=\frac{1}{\hbar^2}\sum_{m,n=1}^2 \langle k|\hat{A}_m|l \rangle \langle l|\hat{A}_n|k \rangle \chi_{mn} S(\omega_{kl}),
\end{equation}
where $\omega_{kl} =E_{kl}/\hbar=E_0(\lambda_k-\lambda_l)/\hbar$. The rates are the off-diagonal elements of $\Gamma_{i\rightarrow j}$ given by 
\begin{eqnarray} \label{ratematrix}
{\bf \Gamma} =\left(
\begin{array}{cccc}
-2\Gamma_{\uparrow}& (1-\chi)\Gamma_{\uparrow} & (1+\chi)\Gamma_{\uparrow} & 0 \\(1-\chi)\Gamma_{\downarrow} & -\Gamma_{\Sigma}(1-\chi) & 0 & (1-\chi)\Gamma_{\uparrow} \\ (1+\chi)\Gamma_{\downarrow} & 0 & -\Gamma_{\Sigma}(1+\chi) & (1+\chi)\Gamma_{\uparrow} \\ 0 & (1-\chi)\Gamma_{\downarrow} & (1+\chi)\Gamma_{\downarrow} & -2\Gamma_{\downarrow}
\end{array}
\right)\nonumber\\
\end{eqnarray}  
where $\Gamma_{\Sigma}=\Gamma_{\uparrow}+\Gamma_{\downarrow}$, $\Gamma_{\downarrow,\uparrow }=\Gamma_{\downarrow,\uparrow ,{\rm C}}+\Gamma_{\downarrow,\uparrow ,{\rm H}}$, and $\Gamma_{\downarrow,\uparrow}=\frac{g^2}{\hbar^2}\frac{\Delta^2}{q^2+\Delta^2}S(\pm \omega_0)$ is the corresponding rate for a single qubit with identical parameters coupled to the noise source, $\hbar\omega_0=2E_0\sqrt{q^2+\Delta^2}$ is the level spacing, and $g$ is the coupling constant \cite{Jukka}.  The effect of $\chi$ on the transition rates between the energy levels for decoupled qubits is illustrated in Fig. \ref{fig2}. According to \eqref{ratematrix} for fully (anti)correlated noise $\chi=+1$ ($\chi=-1$), there are forbidden transitions to/from $|2\rangle$ $(|3\rangle)$, and as a result it becomes a protected state as shown in Fig. \ref{fig2}.

\begin{figure}
\centering
\includegraphics [width=\columnwidth] {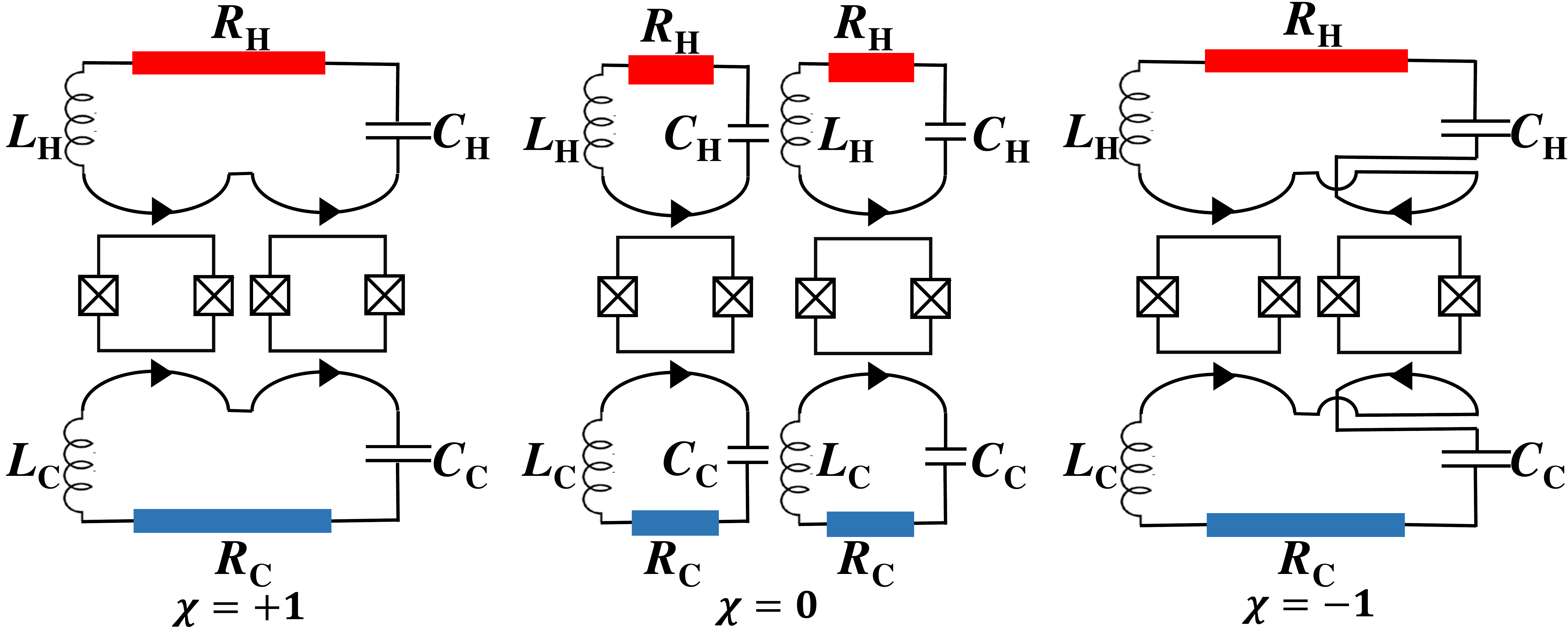}
\caption{Illustration of correlated (${\chi}=+1$, left), uncorrelated ($\chi=0$, center), and anticorrelated ($\chi=-1$, right) noise in the Otto refrigerator configuration.
\label{fig3}}
\end{figure}

To demonstrate the significance of the correlation of noise on measurable quantities, we focus on the system depicted in Fig. \ref{fig3} for different values of $\chi$. We consider noise sources to be thermal baths. This set-up allows us to investigate quantum heat transport between the baths, a topic of considerable experimental interest currently \cite{schwab, jukkanature, jezouin, partanen}. In this case, the fluctuating quantity $\delta \hat X_m(t)$ is presented by electric current noise $\delta i_m(t)$ and the coupling $\hat{A}_m$ is $g_m\sigma_{z,m}$. Here $g_m$ is coupling, e.g., by mutual inductance between each qubit and the fluctuating current. As shown in Fig. \ref{fig3}, the two qubits in the middle are coupled to heat baths at two different temperatures $T_{\rm B}$, where ${\rm B}={\rm C},{\rm H}$ indicates the "cold" and "hot" baths, respectively. The baths are represented by the resistors $R_{\rm B}$ embedded in the $LC$ resonators with a quality factor $Q_{\rm B}=\sqrt{L_{\rm B}/C_{\rm B}}/R_{\rm B}$.

The quantity of interest here is the power transmitted between the hot and cold baths mediated by the qubits. We study the full system by writing the master equation for the density matrix of the system and the environment $\rho_{\rm tot}=\rho \otimes \rho_{\rm E}$ in the interaction picture of the two qubit system as
\begin{equation} \label{dotr}
\dot \rho_{\rm tot} = \frac{i}{\hbar} [\rho_{\rm tot} , H_{\rm D,I}(t)+H_{{\rm cS},I}(t)].
\end{equation}
Here we have assumed that the qubits are driven by time-dependent rotation $H_{\rm D,I}=-i\hbar D^\dagger \dot D$ where $D$ is given by
\begin{eqnarray} \label{Dmatrix}
D =\frac{1}{\sqrt{2}}\left(
\begin{array}{cccc}
1&q/\sqrt{q^2+\Delta^2}&\Delta/\sqrt{q^2+\Delta^2}&0 \\0&0&0&\sqrt{2} \\ 0&\Delta/\sqrt{q^2+\Delta^2}&-q/\sqrt{q^2+\Delta^2}&0\\ 1&-q/\sqrt{q^2+\Delta^2}&-\Delta/\sqrt{q^2+\Delta^2}&0
\end{array}
\right).\nonumber\\
\end{eqnarray}
$H_{{\rm cS},I}$ arises from the noise described above, and presented specifically for similar set-up in \cite{niskanen2007}. The components of $\rho$ can be obtained from the full master equation \cite{breuer} by tracing out the environment with the result
\begin{widetext}
\begin{eqnarray} \label{mefull}
\dot \rho_{kl} = \sum_{i=1}^4 \big {\{}\rho_{ki} \langle i|D^\dagger \dot D|l\rangle e^{i{\Omega}^{-1} \int_0^u \lambda_{il}(u') du'}+ \rho_{il} \langle i|D^\dagger \dot D|k\rangle e^{i{\Omega}^{-1} \int_0^u \lambda_{ki}(u') du'}  +\delta_{kl}\rho_{ii}\Gamma_{i\rightarrow k}-\frac{1}{2}\rho_{kl} (\Gamma_{l\rightarrow i}+\Gamma_{k\rightarrow i})\big {\}}.
\end{eqnarray}
\end{widetext}
Here $\Omega= 2\pi \hbar f/E_0$ denotes the dimensionless frequency of the drive and $u=2\pi ft$ the time, where $f$ is the actual driving frequency.   

The instantaneous power to bath $\rm B$ can then be written as \cite{bayan}
\begin{equation} \label{final.power}
P_{\rm B} = \sum_{k,l} \rho_{kk} E_{kl}\Gamma_{k\rightarrow l,\rm B}. 
\end{equation}  
In the non-driven case the relaxation towards steady state is governed by
\begin{eqnarray}\label{rhos-nondriven}
\dot{\rho}_d={\bf \Gamma}^{\rm T} \rho_d,
\end{eqnarray}
where $\rho_d=(\rho_{11}~~\rho_{22}~~\rho_{33}~~\rho_{44})^{\rm T}$. For $\chi=\pm 1$ the steady-state solution of Eq. (\ref{rhos-nondriven}) depends on the initial condition applied to the system. Due to forbidden transitions according to Eq. (\ref{ratematrix}), the system behaves differently based on its initialization. With $\chi=+1$, if the system is initialized in the state $|2\rangle$ we have $\rho_{22}=1, \rho_{11}=\rho_{33}=\rho_{44}=0$ which demonstrates that $|2\rangle$ is a protected state. On the other hand, initializing in the subspace $\{|1\rangle,|3\rangle,|4\rangle\}$ for $\chi=+1$ leads to $\rho_{22}=0$, $\rho_{11}=\frac{1}{(1+r)^2}$, $\rho_{33}=\frac{r}{(1+r)^2}$, $\rho_{44}=\frac{r^2}{(1+r)^2}$, where $r=\Gamma_{\uparrow}/\Gamma_{\downarrow}$. For $\chi=-1$, one should simply swap states $|2\rangle$ and $|3\rangle$ above. Generally for $\chi\neq \pm 1$ we have
\begin{eqnarray}
&&\rho_{11}=\frac{1}{(1+r)^2},~~
\rho_{22}=\rho_{33}=\frac{r}{(1+r)^2},~~
\rho_{44}=\frac{r^2}{(1+r)^2},\nonumber\\ 
\end{eqnarray} 
which is independent of the correlation $\chi$. Then the steady state power to bath ${\rm C}$ reads
\begin{eqnarray} \label{pcc}
&&P_{\rm C}=\big{[}-(\Gamma_{1\rightarrow 2,{\rm C}}+\Gamma_{1\rightarrow 3,{\rm C}})\rho_{11} +(\Gamma_{2\rightarrow 1,{\rm C}}-\Gamma_{2\rightarrow 4,{\rm C}})\rho_{22}\nonumber \\&&  
+(\Gamma_{3\rightarrow 1,{\rm C}}-\Gamma_{3\rightarrow 4,{\rm C}})\rho_{33} +(\Gamma_{4\rightarrow 2,{\rm C}}+\Gamma_{4\rightarrow 3,{\rm C}})\rho_{44}\big{]}E\nonumber\\&&
 =2(-\rho_{gg}\Gamma_{\uparrow,{\rm C}}+\rho_{ee}\Gamma_{\downarrow,{\rm C}})E = 2P_0,
\end{eqnarray} 
where $\rho_{gg}=1-\rho_{ee}$ is the ground state population of a single qubit in the instantaneous eigenbasis and $P_0$ the transmitted power by it to the cold reservoir \cite{Jukka}.
The power $P_{\rm C}$ {\it in steady state} thus scales with the number of qubits and is independent of the degree of correlation of the noise for $\chi \neq \pm 1$.

\begin{figure}[h]
\centering
\includegraphics [width=8cm] {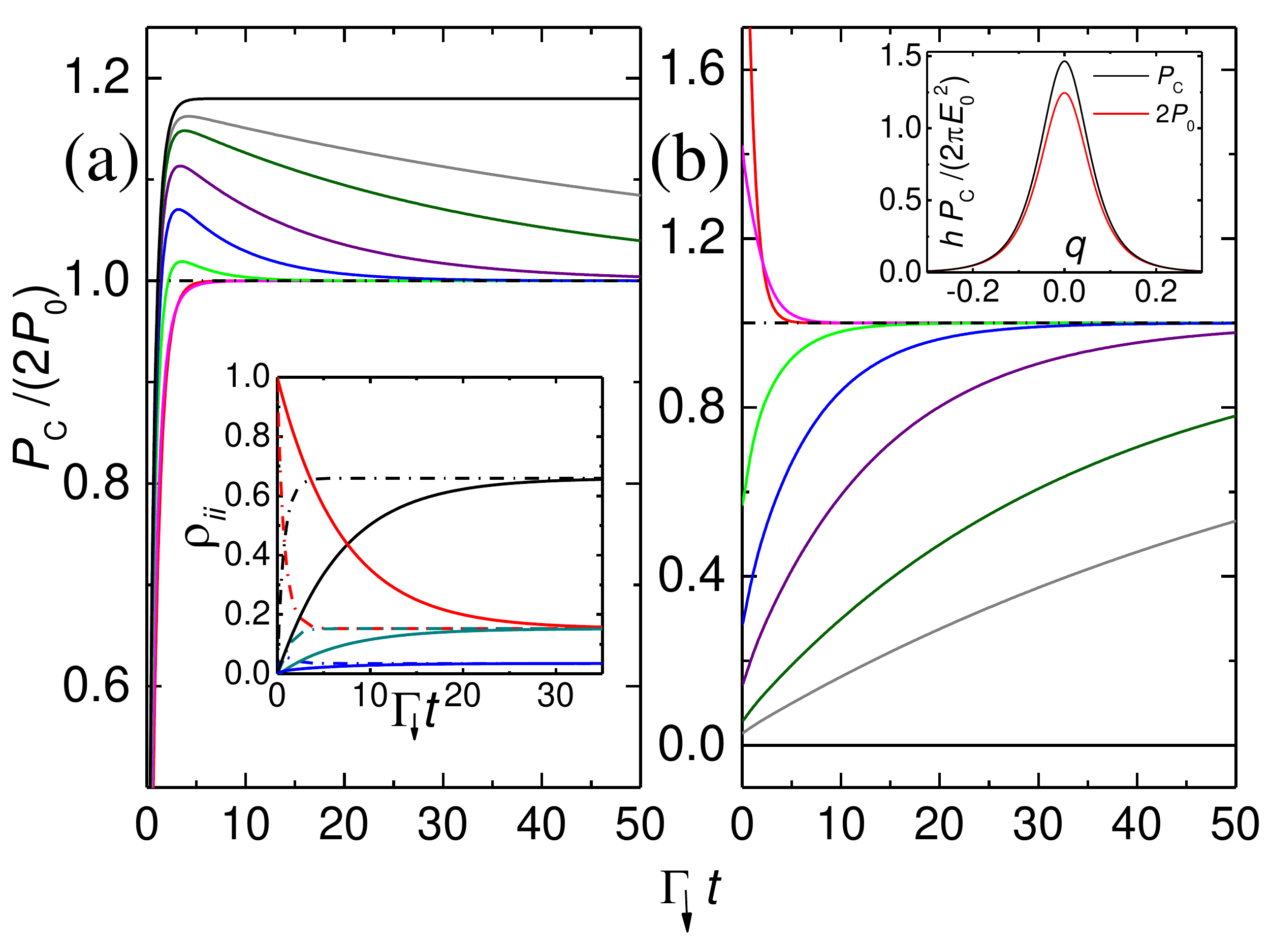}
\caption{The power $P_{\rm C}$ normalized by $2P_0$ (black dot-dashed lines), where $P_0$ is the power of a single qubit, at $q=0$ via the non-driven qubits to the cold bath(s) as a function of time $(\Gamma_{\downarrow}t)$ for various degrees of correlation $\chi$. (a) $\chi=0$, $0.5$, $0.8$, $0.9$, $0.95$, $0.98$, $0.99$, and $1$ from bottom to top; the system is initialized in $|1\rangle$ at $t=0$. (b) The same values and colours for $\chi$ as in (a), initialized in $|2\rangle$  at $t=0$. Inset in (a): Populations $\rho_{11}$ (black lines), $\rho_{22}$ (red lines), $\rho_{33}$ (dark cyan lines), and $\rho_{44}$ (blue lines) when the system is initialized in $|2\rangle$ at $t=0$ for $\chi=0$ (dot-dashed lines) and $\chi=0.9$ (solid lines). Inset in (b): Dimensionless steady state power via the non-driven qubits to the cold bath(s) as a function of detuning $q$ for the case where the noise is fully (anti)correlated (black line) and for other degrees of correlation (red line). The parameters are $\hbar \omega_1/E_0 = \hbar \omega_2/E_0 = 0.1$, $g_1=g_2=1.0$, $k_BT_{\rm H}/E_0 = 0.2$, $k_BT_{\rm C}/E_0 = 0.05$, $Q_{\rm C}=Q_{\rm H}=10$, and $\Delta = 0.1$.
\label{fig4}}
\end{figure} 
\begin{figure}
\centering
\includegraphics [width=\columnwidth] {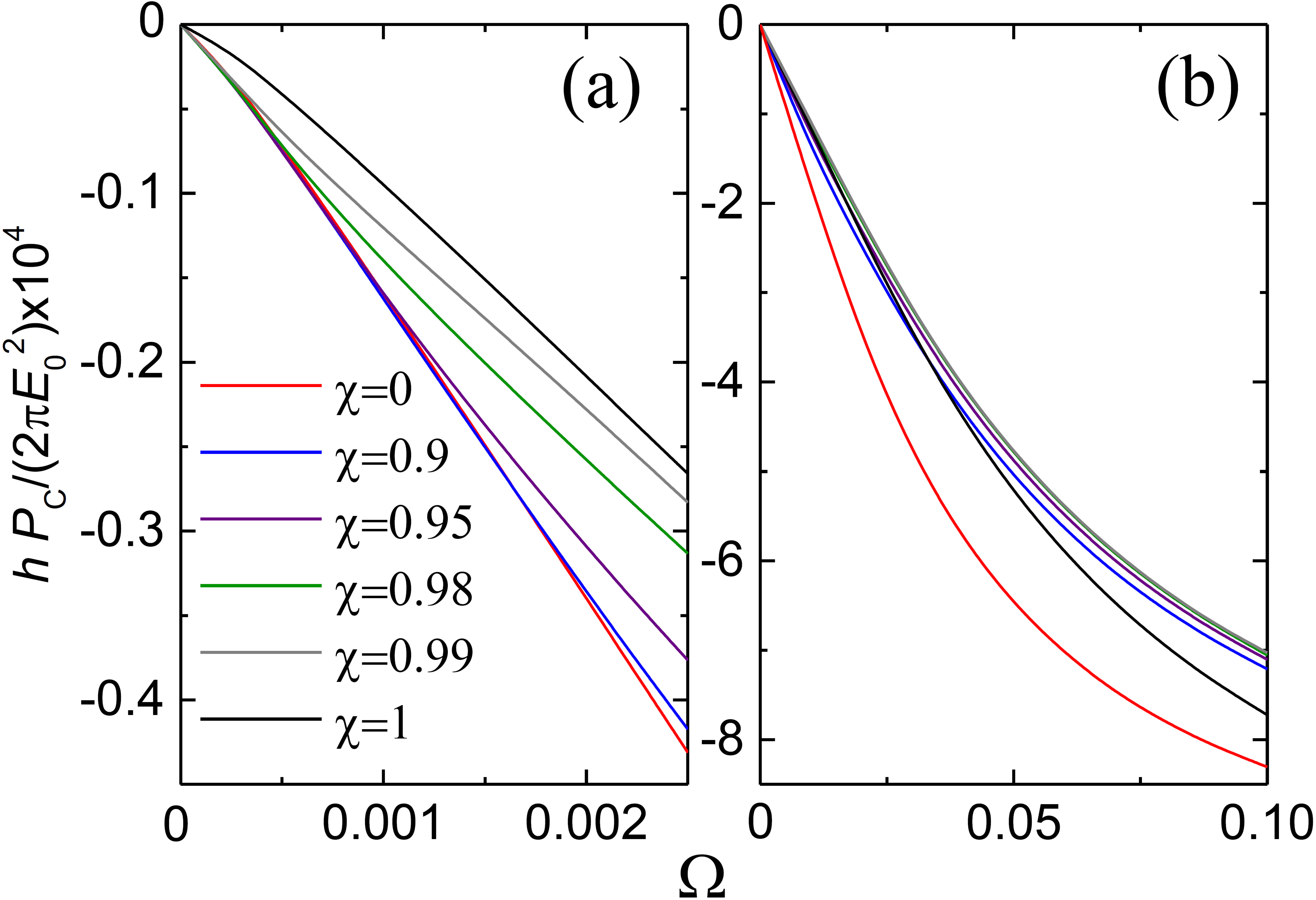}
\caption{The power to the cold bath(s) $P_{\rm C}$ of the two-qubit Otto-refrigerator at (a) low and (b) high frequencies driven by a sinusoidal in time field as a function of frequency for different degrees of correlation $\chi$ indicated within the figure. The system is initialized in the ground state. The parameters are $\hbar \omega_1/E_0 = 2 \sqrt{1/4 + \Delta^2}$, $\hbar \omega_2/E_0 = 2\Delta$, $g_1=g_2=0.25$, $k_BT_{\rm H}/E_0 = k_BT_{\rm C}/E_0 = 0.3$, $Q_{\rm C}=Q_{\rm H}=10$, and $\Delta = 0.3$.
\label{fig5}}
\end{figure}
The inset of Fig. \ref{fig4}b demonstrates the puzzling result that this power is larger than $2P_0$ for the special values $\chi=\pm 1$, i.e. for fully correlated or anticorrelated noise. The origin of this result becomes obvious by looking at the dynamics of the density matrix after the system has been initialized in an arbitrary state. Due to the presence of a protected state $|2\rangle$ ($|3\rangle$) for $\chi=+1$ ($\chi=-1$), the power of Eq. \eqref{pcc} cannot be reached in finite time for the case of fully (anti)correlated noise. The time dependence of power $P_{\rm C}$ normalized by the uncorrelated power $2P_0$ when the system is initialized in state $|1\rangle$ and $|2\rangle$, respectively, at $t=0$ are shown in Fig. \ref{fig4}a,b. The asymptotic value of $P_{\rm C}$ does not depend on this initial state, except for the cases $\chi=\pm 1$. It is seen here that for $\chi\rightarrow +1$, the time to approach $P_{\rm C} = 2P_0$ becomes longer and longer, and finally, this relaxation time becomes infinite for $\chi=+1$. This explains the result in the inset of Fig. \ref{fig4}b, where power is enhanced for $\chi=\pm 1$ above that of the other values of $\chi$. The same slow relaxation for higher values of $\chi$ is seen in the inset of Fig. \ref{fig4}a, where we plot the populations of the states $\rho_{ii}$ $(i=1,2,3,4)$ when the system is initialized in $|2\rangle$ for two different values of $\chi=0$ and $0.9$. It is also worth mentioning that if one initializes the system to the state $|2\rangle$ for $\chi=+1$ or $|3\rangle$ for $\chi=-1$, the power $P_{\rm C}=0$, as this state is a protected one. For any other value of $\chi$ away from $\chi=\pm 1$ the power approaches $2P_0$ after a sufficiently long time also in this case.

We discuss finally a quantum Otto refrigerator \cite{Jukka, niskanen2007}. In this device, applying periodic time dependent drive to the qubits in Fig. \ref{fig3}, heat can be transferred from the cold bath to the hot one, provided $\omega_{\rm H}=1/\sqrt{L_{\rm H}C_{\rm H}}>\omega_{\rm C}=1/\sqrt{L_{\rm C}C_{\rm C}}$. We introduce a standard driving protocol $q(t)=(1+\cos 2\pi ft)/4$. In the numerical results, the power is averaged over one cycle once it has reached the  steady state. Solving the general master equation \eqref{mefull} numerically, we plot the frequency dependence of the cooling power of the quantum refrigerator for different degrees of correlation in two different frequency ranges in Fig. \ref{fig5}a,b. It is vivid that at very low frequencies the curves for all values of $\chi$ (except $\chi=+1$) in Fig. \ref{fig5}a collapse on each other. They start to deviate from the curve at $\chi=0$ at the critical frequency $\Omega_c\propto (1- \chi)$. This is because of the competition between the slowest transition rates $\propto(1-\chi)$ in Eq. \eqref{ratematrix} to/from $|2\rangle$ and the driving frequency $\Omega$. At higher frequencies, $\Omega \gg \Omega_c$, the transitions to/from state $|2\rangle$ cannot follow the drive (Fig. \ref{fig5}b), $|2\rangle$ is thus dynamically protected and we effectively  deal with a three-level system. Thus for $\Omega\gg\Omega_c$ , all the curves with $\chi\sim +1$ converge to the same value. In this regime $\rho_{22}$ has a small but essentially time-independent value. For $\chi=+1$, $\rho_{22}=0$ as the system was initialized in state $|1\rangle$. For this particular value of $\chi$ the power $-P_{\rm C}$ is again higher than in the partially correlated case.

In conclusion, we have investigated the golden-rule transition rates between the four energy levels of a two-qubit system when it is subjected to fully and partially (anti)correlated noise sources. By tuning the degree of correlation of the noise sources, we demonstrate protected states and variations in the transmitted power between thermal baths. This power exhibits a different steady-state value in the presence of a protected state as opposed to that of the standard four level system. In particular, the former power vanishes when the qubits are initialized in a protected state. Moreover, for nearly (anti)correlated noise, there is a long relaxation time to reach the steady state level of power which is fully independent of the level of correlation of the noise for $\chi\neq \pm 1$. Under AC driven conditions, there is an interesting interplay between this slow relaxation rate and the driving frequency, which governs the power of a quantum refrigerator that we present as an example.  

We acknowledge Yuri Galperin, Rosario Fazio and Michele Campisi for useful discussions. The work was supported by the Academy of Finland (grants 272218 and 284594).


\begin{thebibliography}{99}
\bibitem{clarke} J. Clarke and F. K. Wilhelm, Superconducting quantum bits, Nature {\bf 453}, 1031 (2008).

\bibitem{devoret} M. H. Devoret and R. J. Schoelkopf, Superconducting circuits for quantum information: An outlook, Science {\bf 339}, 1169 (2013).

\bibitem{nielsen} Michael A. Nielsen and Isaac L. Chuang, {\it Quantum computation and quantum information}, (Cambridge University Press, Cambridge, 2010).

\bibitem{governale} Michele Governale, Milena Grifoni, Gerd Sch\"on, Decoherence and dephasing in coupled Josephson-junction qubits, Chemical Physics {\bf 268}, 273 (2001).                          

\bibitem{storcz} Markus J. Storcz and Frank K. Wilhelm, Decoherence and gate performance of coupled solid-state qubits, Phys. Rev. A {\bf 67}, 042319 (2003).                          

\bibitem{pashkin} Yu. A. Pashkin, T. Yamamoto, O. Astafiev, Y. Nakamura, D. V. Averin, J. S. Tsai, Quantum oscillations in two coupled charge qubits, Nature {\bf 42}, 823 (2003).

\bibitem{averin} Kristian Rabenstein and Dmitri V. Averin, Decoherence in two coupled qubits, Turk J. Phys. {\bf 27}, 313 (2003). 

\bibitem{you} J. Q. You, Xuedong Hu, and Franco Nori, Correlation-induced suppression of decoherence in capacitively coupled Cooper-pair boxes, Phys. Rev. B {\bf 72}, 144529 (2005).   

\bibitem{hu} Yong Hu, Zheng-Wei Zhou, Jian-Ming Cai, and Guang-Can Guo, Decoherence of coupled Josephson charge qubits due to partially correlated low-frequency noise, Phys. Rev. A {\bf 75}, 052327 (2007).                       


\bibitem{zanardi} P. Zanardi and M. Rasetti, Noiseless quantum codes, Phys. Rev. Lett. {\bf 79}, 3306 (1997).                       

\bibitem{d'arrigo} A. D'Arrigo, A. Mastellone, E. Paladino, and G. Falci, Effects of low-frequency noise cross-correlations in coupled superconducting qubits, New Journal of Physics {\bf 10}, 115006 (2008).                       

\bibitem{gustavsson} S. Gustavsson, J. Bylander, F. Yan, W. D. Oliver, F. Yoshihara, and Y. Nakamura, Noise correlations in a flux qubit with tunable tunnel coupling, Phys. Rev. B {\bf 84}, 014525 (2011).

\bibitem{galperin} H. Brox, J. Bergli, and Y. M. Galperin, Bloch-sphere approach to correlated noise in coupled qubits, J. Phys. A: Math. Theor. {\bf 45}, 455302 (2012). 

\bibitem{barends} R. Barends {\it et al.}, Superconducting quantum circuits at the surface code threshold for fault tolerance, Nature {\bf 508}, 500 (2014).

\bibitem{averin1} D. V. Averin and R. Fazio, Active suppression of dephasing in Josephson-junction qubits, JETP Lett. {\bf 78}, 1162 (2003).

\bibitem{clemens} James P. Clemens, Shabnam Siddiqui, and Julio Gea-Banacloche, Quantum error correction against correlated noise, Phys. Rev. A {\bf 69}, 062313 (2004).

\bibitem{klesse} Rochus Klesse and Sandra Frank, Quantum error correction in spatially correlated quantum noise, Phys. Rev. Lett. {\bf 95}, 230503 (2005).

\bibitem{novais} E. Novais and Harold U. Baranger, Decoherence by correlated noise and quantum error correction, Phys. Rev. Lett. {\bf 97}, 040501 (2006).
                    
\bibitem{Jukka} B. Karimi, J. P. Pekola, Otto refrigerator based on a superconducting qubit: Classical and quantum performance, Phys. Rev. B {\bf 94}, 184503 (2016).

\bibitem{niskanen2007} A. O. Niskanen, Y. Nakamura, and J. P. Pekola, Information entropic superconducting microcooler, Phys. Rev. B {\bf 76}, 174523 (2007).


\bibitem{kay} Kay Brandner, Michael Bauer, and Udo Seifert, Universal coherence-induced power losses of quantum heat engines in linear response, arXiv:1703.02464.

\bibitem{campisi2016} M. Campisi and R. Fazio, The power of a critical heat engine, Nat. Commun. {\bf 7}, 11895 (2016).

\bibitem{hofer2016} P. P. Hofer, J.-R. Souquet, and A. A. Clerk, Quantum heat engine based on photon-assisted Cooper pair tunneling, Phys. Rev. B {\bf 93}, 041418(R) (2016).


\bibitem{quan2007} H. T. Quan, Yu-xi Liu, C. P. Sun, and F. Nori, Quantum thermodynamic cycles and quantum heat engines, Phys. Rev. E {\bf 76}, 031105 (2007).

\bibitem{rossnagel2016} J. Rossnagel, S. T. Dawkins, K. N. Tolazzi, O. Abah, E. Lutz, F. Schmidt-Kaler, and K. Singer, A single-atom heat engine, Science {\bf 352}, 6283 (2016).

\bibitem{Uzdin/Kosloff} R. Uzdin, A. Levy, and R. Kosloff, Equivalence of quantum heat machines, and quantum-thermodynamic signatures, Physical Review X {\bf 5}, 031044 (2015).

\bibitem{brandner2016} K. Brandner and U. Seifert, Periodic thermodynamics of open quantum systems, Phys. Rev. E {\bf 93}, 062134 (2016).

\bibitem{breuer} H.-P. Breuer and F. Petruccione,{\it The theory of open quantum systems} (Oxford University Press, Oxford, 2002).

\bibitem{schwab} K. Schwab, E. A. Henriksen, J. M. Worlock, and M. L. Roukes, Measurement of the quantum of thermal conductance, Nature {\bf 404}, 974 (2000).

\bibitem{jukkanature} Matthias Meschke, Wiebke Guichard, and Jukka P. Pekola, Single-mode heat conduction by photons, Nature {\bf 444}, 05276 (2006).

\bibitem{jezouin} S. Jezouin, F. D. Parmentier, A. Anthore, U. Gennser, A. Cavanna, Y. Jin, F. Pierre, Quantum limit of heat flow across a single electronic channel, Science {\bf 342}, 601-604 (2013).

\bibitem{partanen} Matti Partanen, Kuan Yen Tan, Joonas Govenius, Russell E. Lake, Miika K. M\"akel\"a, Tuomo Tanttu and Mikko M\"ott\"onen, Quantum-limited heat conduction over macroscopic distances, Nature Physics  {\bf 12},  460 (2016).

\bibitem{bayan} B. Karimi, J. P. Pekola, M. Campisi, and R. Fazio, Coupled qubits as a quantum heat switch, submitted (2017).

\end{thebibliography}
\end{document}